

%
\newbox\leftpage \newdimen\fullhsize \newdimen\hstitle \newdimen\hsbody
\tolerance=1000\hfuzz=2pt
\def\bigans{l }
%
\ifx\answ\bigans\message{(This will come out unreduced.}
\magnification=1000\baselineskip=18pt plus 2pt minus 1pt
\hsbody=\hsize \hstitle=\hsize 
\else\def\apans{l }
\let\lr=L
\magnification=1000\baselineskip=16pt plus 2pt minus 1pt
\voffset=-.31truein\vsize=7truein
\hstitle=8truein\hsbody=4.75truein\fullhsize=10truein\hsize=\hsbody
\ifx\apansw\apans\special{ps: landscape}\hoffset=-.59truein
  \else\hoffset=.05truein\fi
\output={\ifnum\pageno=0 
  \shipout\vbox{\hbox to \fullhsize{\hfill\pagebody\hfill}}\advancepageno
  \else
  \almostshipout{\leftline{\vbox{\pagebody\makefootline}}}\advancepageno
  \fi}
\def\almostshipout#1{\if L\lr \count1=1
      \global\setbox\leftpage=#1 \global\let\lr=R
  \else \count1=2
    \shipout\vbox{\ifx\apansw\apans\special{ps: landscape}\fi 
      \hbox to\fullhsize{\box\leftpage\hfil#1}}  \global\let\lr=L\fi}
\fi
%
\catcode`\@=11 
\newcount\yearltd\yearltd=\year\advance\yearltd by -1900

\def\Title#1#2{\nopagenumbers\abstractfont\hsize=\hstitle\rightline{#1}%
\vskip 1in\centerline{\titlefont #2}\abstractfont\vskip .5in\pageno=0}
\def\Date#1{\vfill\leftline{#1}\tenpoint\supereject\global\hsize=\hsbody%
\footline={\hss\tenrm\folio\hss}}
\def\draftmode{\def\draftdate{{\rm preliminary draft:
\number\month/\number\day/\number\yearltd\ \ \hourmin}}%
\headline={\hfil\draftdate}\writelabels\baselineskip=20pt plus 2pt minus 2pt
{\count255=\time\divide\count255 by 60 \xdef\hourmin{\number\count255}
	\multiply\count255 by-60\advance\count255 by\time
   \xdef\hourmin{\hourmin:\ifnum\count255<10 0\fi\the\count255}}}

\def\nolabels{\def\eqnlabel##1{}\def\eqlabel##1{}\def\reflabel##1{}}
\def\writelabels{\def\eqnlabel##1{%
{\escapechar=` \hfill\rlap{\hskip.09in\string##1}}}%
\def\eqlabel##1{{\escapechar=` \rlap{\hskip.09in\string##1}}}%
\def\reflabel##1{\noexpand\llap{\string\string\string##1\hskip.31in}}}
\nolabels
%
\global\newcount\secno \global\secno=0
\global\newcount\meqno \global\meqno=1
\def\newsec#1{\global\advance\secno by1
\xdef\secsym{\the\secno.}\global\meqno=1
\bigbreak\bigskip
\noindent{\bf\the\secno. #1}\par\nobreak\medskip\nobreak}
\xdef\secsym{}
\def\appendix#1#2{\global\meqno=1\xdef\secsym{\hbox{#1.}}\bigbreak\bigskip
\noindent{\bf Appendix #1. #2}\par\nobreak\medskip\nobreak}
%
%
\def\eqnn#1{\xdef #1{(\secsym\the\meqno)}%
\global\advance\meqno by1\eqnlabel#1}
\def\eqna#1{\xdef #1##1{\hbox{$(\secsym\the\meqno##1)$}}%
\global\advance\meqno by1\eqnlabel{#1$\{\}$}}
\def\eqn#1#2{\xdef #1{(\secsym\the\meqno)}\global\advance\meqno by1%
$$#2\eqno#1\eqlabel#1$$}
%
\newskip\footskip\footskip14pt plus 1pt minus 1pt 
\def\f@@t{\baselineskip\footskip\bgroup\aftergroup\@foot\let\next}
\setbox\strutbox=\hbox{\vrule height9.5pt depth4.5pt width0pt}
\global\newcount\ftno \global\ftno=0
\def\foot{\global\advance\ftno by1\footnote{$^{\the\ftno}$}}
%
%
\global\newcount\refno \global\refno=1
\newwrite\rfile
\def\ref{[\the\refno]\nref}
\def\nref#1{\xdef#1{[\the\refno]}\ifnum\refno=1\immediate
\openout\rfile=refs.tmp\fi\global\advance\refno by1\chardef\wfile=\rfile
\immediate\write\rfile{\noexpand\item{#1\ }\reflabel{#1}\pctsign}\findarg}
\def\findarg#1#{\begingroup\obeylines\newlinechar=`\^^M\pass@rg}
{\obeylines\gdef\pass@rg#1{\writ@line\relax #1^^M\hbox{}^^M}%
\gdef\writ@line#1^^M{\expandafter\toks0\expandafter{\striprel@x #1}%
\edef\next{\the\toks0}\ifx\next\em@rk\let\next=\endgroup\else\ifx\next\empty%
\else\immediate\write\wfile{\the\toks0}\fi\let\next=\writ@line\fi\next\relax}}
\def\striprel@x#1{} \def\em@rk{\hbox{}} {\catcode`\%=12\xdef\pctsign{
\def\semi{;\hfil\break}
\def\addref#1{\immediate\write\rfile{\noexpand\item{}#1}} 
\def\listrefs{\vfill\eject\immediate\closeout\rfile
\baselineskip=12pt\centerline{{\bf References}}\bigskip{\frenchspacing%
\escapechar=` \input refs.tmp\vfill\eject}\nonfrenchspacing}
\def\startrefs#1{\immediate\openout\rfile=refs.tmp\refno=#1}
\def\figures{\centerline{{\bf Figure Captions}}\medskip\parindent=40pt}
\def\fig#1#2{\medskip\item{Figure ~#1:  }#2}
\catcode`\@=12 
%
\ifx\answ\bigans
\font\titlerm=cmr10 scaled\magstep3 \font\titlerms=cmr7 scaled\magstep3
\font\titlermss=cmr5 scaled\magstep3 \font\titlei=cmmi10 scaled\magstep3
\font\titleis=cmmi7 scaled\magstep3 \font\titleiss=cmmi5 scaled\magstep3
\font\titlesy=cmsy10 scaled\magstep3 \font\titlesys=cmsy7 scaled\magstep3
\font\titlesyss=cmsy5 scaled\magstep3 \font\titleit=cmti10 scaled\magstep3
\else
\font\titlerm=cmr10 scaled\magstep4 \font\titlerms=cmr7 scaled\magstep4
\font\titlermss=cmr5 scaled\magstep4 \font\titlei=cmmi10 scaled\magstep4
\font\titleis=cmmi7 scaled\magstep4 \font\titleiss=cmmi5 scaled\magstep4
\font\titlesy=cmsy10 scaled\magstep4 \font\titlesys=cmsy7 scaled\magstep4
\font\titlesyss=cmsy5 scaled\magstep4 \font\titleit=cmti10 scaled\magstep4
\font\absrm=cmr10 scaled\magstep1 \font\absrms=cmr7 scaled\magstep1
\font\absrmss=cmr5 scaled\magstep1 \font\absi=cmmi10 scaled\magstep1
\font\absis=cmmi7 scaled\magstep1 \font\absiss=cmmi5 scaled\magstep1
\font\abssy=cmsy10 scaled\magstep1 \font\abssys=cmsy7 scaled\magstep1
\font\abssyss=cmsy5 scaled\magstep1 \font\absbf=cmbx10 scaled\magstep1
\skewchar\absi='177 \skewchar\absis='177 \skewchar\absiss='177
\skewchar\abssy='60 \skewchar\abssys='60 \skewchar\abssyss='60
\fi
\skewchar\titlei='177 \skewchar\titleis='177 \skewchar\titleiss='177
\skewchar\titlesy='60 \skewchar\titlesys='60 \skewchar\titlesyss='60
\def\titlefont{\def\rm{\fam0\titlerm}
\textfont0=\titlerm \scriptfont0=\titlerms \scriptscriptfont0=\titlermss
\textfont1=\titlei \scriptfont1=\titleis \scriptscriptfont1=\titleiss
\textfont2=\titlesy \scriptfont2=\titlesys \scriptscriptfont2=\titlesyss
\textfont\itfam=\titleit \def\it{\fam\itfam\titleit} \rm}
\ifx\answ\bigans\def\abstractfont{\tenpoint}\else
\def\abstractfont{\def\rm{\fam0\absrm}
\textfont0=\absrm \scriptfont0=\absrms \scriptscriptfont0=\absrmss
\textfont1=\absi \scriptfont1=\absis \scriptscriptfont1=\absiss
\textfont2=\abssy \scriptfont2=\abssys \scriptscriptfont2=\abssyss
\textfont\itfam=\bigit \def\it{\fam\itfam\bigit}
\textfont\bffam=\absbf \def\bf{\fam\bffam\absbf} \rm} \fi
\def\tenpoint{\def\rm{\fam0\tenrm}
\textfont0=\tenrm \scriptfont0=\sevenrm \scriptscriptfont0=\fiverm
\textfont1=\teni  \scriptfont1=\seveni  \scriptscriptfont1=\fivei
\textfont2=\tensy \scriptfont2=\sevensy \scriptscriptfont2=\fivesy
\textfont\itfam=\tenit \def\it{\fam\itfam\tenit}
\textfont\bffam=\tenbf \def\bf{\fam\bffam\tenbf} \rm}
%
%
\def\noblackbox{\overfullrule=0pt}
\hyphenation{anom-aly anom-alies coun-ter-term coun-ter-terms}
\def\inv{^{\raise.15ex\hbox{${\scriptscriptstyle -}$}\kern-.05em 1}}
\def\dup{^{\vphantom{1}}}
\def\Dsl{\,\raise.15ex\hbox{/}\mkern-13.5mu D} 
\def\dsl{\raise.15ex\hbox{/}\kern-.57em\partial}
\def\del{\partial}
\def\Psl{\dsl}
\def\tr{{\rm tr}} \def\Tr{{\rm Tr}}
\font\bigit=cmti10 scaled \magstep1
\def\biglie{\hbox{\bigit\$}} 
\def\lspace{\ifx\answ\bigans{}\else\qquad\fi}
\def\lbspace{\ifx\answ\bigans{}\else\hskip-.2in\fi} 
\def\boxeqn#1{\vcenter{\vbox{\hrule\hbox{\vrule\kern3pt\vbox{\kern3pt
	\hbox{${\displaystyle #1}$}\kern3pt}\kern3pt\vrule}\hrule}}}
\def\mbox#1#2{\vcenter{\hrule \hbox{\vrule height#2in
		\kern#1in \vrule} \hrule}}  
%
\def\CAG{{\cal A/\cal G}}   
\def\CA{{\cal A}} \def\CC{{\cal C}} \def\CF{{\cal F}} \def\CG{{\cal G}}
\def\CL{{\cal L}} \def\CH{{\cal H}} \def\CI{{\cal I}} \def\CU{{\cal U}}
\def\CB{{\cal B}} \def\CR{{\cal R}} \def\CD{{\cal D}} \def\CT{{\cal T}}
\def\e#1{{\rm e}^{^{\textstyle#1}}}
\def\grad#1{\,\nabla\!_{{#1}}\,}
\def\gradgrad#1#2{\,\nabla\!_{{#1}}\nabla\!_{{#2}}\,}
\def\ph{\varphi}
\def\psibar{\overline\psi}
\def\om#1#2{\omega^{#1}{}_{#2}}
\def\vev#1{\langle #1 \rangle}
\def\lform{\hbox{$\sqcup$}\llap{\hbox{$\sqcap$}}}
\def\darr#1{\raise1.5ex\hbox{$\leftrightarrow$}\mkern-16.5mu #1}
\def\lie{\hbox{\it\$}} 
\def\ha{{1\over2}}
\def\half{{\textstyle{1\over2}}} 
\def\roughly#1{\raise.3ex\hbox{$#1$\kern-.75em\lower1ex\hbox{$\sim$}}}

\font\names=cmbx10 scaled\magstep1

\def\sg{{\sigma_8({\rm gal})}}
\def\sc{{\sigma_8({\rm CDM})}}
\def\bias{{\sigma_8^{-1}({\rm CDM})}}
\baselineskip=20pt
\Title{PUP-TH-93/1393, hep-ph/9305261}
{\vbox{\centerline
{Degree Scale Microwave Anisotropies}\centerline{ in}
\centerline{
NonGaussian Theories of Cosmic Structure Formation}}}
\font\large=cmr10 scaled\magstep3
\font\names=cmbx10 scaled\magstep1
\centerline{ \bf\names David Coulson$^{1,3}$}
\centerline{\bf\names Ue-Li Pen$^2$}
\centerline{ and }
\centerline{\bf\names Neil Turok$^3$}

\centerline{$^1$ The Blackett Laboratory, Imperial College, London SW7, UK. }

\centerline{$^2$ Astrophysical Sciences, Peyton Hall, Princeton University,
Princeton NJ08544.
 }
\centerline{$^3$ Joseph Henry Laboratories,  Princeton University, Princeton,
NJ08544.}

\centerline{\bf Abstract}
\baselineskip=12pt

The COBE satellite's
maps of the cosmic microwave background (CMB) anisotropy
do not have the resolution to discriminate  between
theories of cosmic structure formation based on
inflation, and those based on field ordering
following a symmetry breaking phase transition. For this purpose
it is critical to resolve the CMB anisotropy on degree scales.
In this paper we report on detailed calculations of the
degree scale anisotropies predicted in the global string, monopole,
texture and nontopological texture scenarios of structure formation,
emphasising their distinct character from those predicted
by inflation, and commenting on the prospects of their detection in
the near future.

\Date{5/93} 
\baselineskip=24pt
\centerline{\bf I. Introduction}

The COBE satellite's detection of large angular scale anisotropy  in
the cosmic microwave background
 \ref\COBE{G. F. Smoot {\it et. al}, Ap. J. {\bf 396}, L1
(1992).} has opened a new window on the early universe.
 But the picture COBE itself provides is
limited by a large angular smoothing
 scale and a low signal to noise ratio.
It falls short of discriminating between
a large number of  theories of structure formation,
based on very different physics. For example, theories based on
inflation and on cosmic field ordering
are equally consistent with COBE's findings.
Both sets of theories predict an approximately scale invariant
spectrum
of multipole moments, and on COBE scales a very Gaussian anisotropy
pattern.
In particular,  the distinctive nonGaussian
signatures expected in the field ordering theories (lines
for strings \ref\ks{N. Kaiser and A. Stebbins, Nature
{\bf 310}, 391-393 (1984).}, spots for textures \ref\ts{N. Turok
and D. Spergel, Phys. Rev. Lett. {\bf 64}, 2736 (1990).}) would
be smeared out by the COBE beam
and instrument noise, and very hard to distinguish from
the Gaussian noise pattern expected from inflation \ref\Ben{
D. P. Bennett and S. H. Rhie, Livermore preprint (1992).}
\ref\PST{U. Pen, D. Spergel and N. Turok,
Princeton preprint (1993).}. For this purpose,
 resolution on an angular scale smaller
than that subtended by the horizon at last scattering appears
to be  essential.

{\it None} of the preexisting  theories  can claim the COBE findings
as an unmitigated success. Roughly speaking,  the COBE result
was a factor of two higher than expected in the
simplest inflationary theory, and a factor of two lower than
expected in the simplest field ordering theories. These
theories have $\Omega=1, \Lambda=0$, with
one free parameter determining the
level of fluctuations - if this is fixed from COBE,
the `standard' inflation plus cold dark matter (CDM) theory predicts
galaxy cluster velocity dispersions much higher than observed
\ref\WEF{S.D.M. White, G. Efstathiou and C.S. Frenk, M.N.R.A.S.,
to appear (1993).}.
It is fairly easy to fix the
situation by adding new adjustable parameters.
In the inflationary case, considering `mixed' dark matter
\ref\MDM{M. Davis, F.J. Summers and D. Schlegel,  Nature, {\bf 359}
(1992)
393. }, inflationary potentials which yield an important
additional
gravity wave  component
 \ref\KW{L. Krauss and M. White, Phys. Rev. Lett.
{\bf 69} (1992) 869; R.L. Davis, {\it et. al.}, Phys. Rev. Lett.
{\bf 69} (1992) 1856,},
\ref\RD{R. Crittenden {\it et. al.} U. Penn. preprint
(1993).},
or adding a cosmological constant \ref\CO{R.
Cen, N. Y. Gnedin and J.P. Ostriker, Princeton preprint 1993.},
all improve the situation.

The simplest field ordering theories are in principle as predictive
as the simplest inflationary scenario, since the level
of mass fluctuations and of microwave anisotropies depend
on a single free parameter, the symmetry breaking scale $\phi_0$.
An appealing aspect of these scenarios is that one naturally
produces the correct amplitude of density fluctuations
with  $\phi_0$  of order the grand unification
scale, around $10^{16}$ GeV. But unlike the simplest inflationary
theory, these theories rely on  nonlinear effects in an important
way,
and are
harder
to calculate with. The most detailed calculations so far
suggest that if the parameter $\phi_0$ is normalised to fit
the COBE $10^o$ variance,
the level of mass fluctuations inferred on
intermediate scales ($\sim 20 h^{-1}$ Mpc) is too small
to explain the observed galaxy streaming motions \PST.
It has been suggested that this problem would be ameliorated
in an open universe
\ref\Sper{D. Spergel, Princeton preprint 1993. }, by suppressing
the anisotropies produced by the field ordering at late times
and thus allowing a larger value for $\phi_0$.
A positive cosmological constant could be expected to have the
same effect.

The pattern of CMB anisotropy on degree scales  offers an
independent and in principle
far cleaner test, being unaffected by the complex processes involved
in galaxy formation.
The simplest measure of anisotropy, and the
only relevant quantity in Gaussian theories,  is the power spectrum of
multipole moments.
If the standard inflation
plus CDM theory is correct, there should be a peak in the spectrum
at $l\sim 200$ produced at the epoch of recombination, $Z\sim 1000$,
due in part to the Doppler effect of photons scattering off moving
electrons. In the theories considered here which involve
topological defects ($N=2,3,4$), reionisation is likely,
and last scattering would be shifted forward to $Z\sim 100$.
The angular scale subtended by the horizon at last scattering
is (for $\Omega=1$, $\Lambda=0$) $\theta_{ls}= (1+Z_{ls})^{-{1\over2}}$,
of order $2^o$ without reionisation and $6^o$ with reionisation.
As will be seen, the spectrum predicted by the field ordering
theories with the assumption of reionisation, is quite
distinct, with much less power in higher multipoles.

Even more distinctive is
the nonGaussianity in the field ordering theories.
The assumption that the anisotropy pattern is in the form
of random Gaussian noise is frequently made. However,
if the microwave sky  is nonGaussian,
there will be additional to be gleaned from
anisotropy maps.
In the nonGaussian theories explored here for example, details of the
pattern would
tell us about the
pattern of symmetry breaking
in particle physics at very high energies, information
presently available
from no other source (see e.g. \ref\JT{M. Joyce and N. Turok,
Princeton preprint 1993.}).

The data gathered so far  from experiments at the
South pole present an incomplete but nonetheless intriguing picture.
If the the  upper limit $(\delta T/T)< 1.4
\times 10^{-6}$ derived from one of the four channels
of the Gaier et. al. experiment \ref\Gai{T. Gaier, J. Schuster,
J. Gundersen, T. Koch, M. Seiffert, P. Meinhold and P. Lubin,
Ap. J. {\bf 398}, L1 (1992).} is correct, it
is enough to exclude
the standard inflationary spectrum of fluctuations (normalised
to COBE) with
either hot or cold dark matter at the 95 \% confidence level
 \ref\Gor{K. Gorski, R. Stompor and R. Juskiewicz,
CNRS preprint  (1992).}.  And the thirteen point scan from the Schuster et. al.
experiment \ref\Sch{J. Schuster, {\it et. al} Santa Barbara preprint
(1993).}
contains an apparently nonGaussian `bump'.
Similarly, data from the MAX experiment seem
inconsistent with
with Gaussian statistics. Under that assumption
discrepant
upper and lower 95 \% confidence bounds were obtained on the level
of CMB anisotropy at ${1\over 2}^o$ from two different parts of the sky
 \ref\Gun{J. O.
Gundersen et. al. CfPA preprint, Berkeley, 1993;
P. Meinhold et. al. CfPA preprint, Berkeley, 1993.}.
It is clearly of  interest to determine the predictions
of the nonGaussian field ordering theories on these angular scales,
while the observational situation is still being resolved.

One of the sharpest distinctions between
the Gaussian and nonGaussian  scenarios mentioned above
is the likelihood that the universe was reionised after recombination.
If around $0.1-1.0$ per cent of the mass in the universe
underwent nonlinear gravitational clustering and star formation
at redshifts greater than of the order 100, this would likely
release sufficient ionising
radiation to
reionise the intergalactic medium, causing the photons to
rescatter.
In this case the epoch of last scattering would be shifted forward
to a redshift  $Z_{ls} \sim 100 (.05/f \Omega_B h_{50})^{2\over 3}$
where $f$ is the fractional ionisation (assumed not too different
from unity).
In field ordering theories with
topological defects like strings or textures, large local density
perturbations are produced at high redshift, and complete
reionisation
($f=1$) is possible
 \PST. But in Gaussian scale invariant theories
of structure formation, reionisation is very unlikely because
high amplitude fluctuations are exponentially suppressed -
using the Press-Schechter \ref\PS{
W. H. Press and P.L. Schechter, Ap. J. {\bf 187} (1974) 452. } or
number of peaks formulae \ref\Bet{J. M.  Bardeen, J. R. Bond,
N. Kaiser and A. Szalay, Ap. J. {\bf 304} (1984) 15.}
one deduces that for the standard CDM theory
normalised to COBE, no more than of order $10^{-10}$ of the baryons
in the universe could have undergone collapse onto dark matter lumps,
and thus be bound into stars, by a redshift of 100.
This is far too little to reionise
the universe.

In this paper we shall make the assumption that
the universe was reionised at redshifts greater than
$Z_{ls}$ given above,
and study the consequences for the pattern of microwave anisotropy.
We shall give results for the simplest symmetry breaking theories:
where an assumed  $O(N)$ global symmetry is spontaneously broken
by an $N$ component scalar field taking a GUT scale vacuum
expectation value.  For
$N=2$, this is the theory of  global cosmic strings,
$N=3$, global monopoles, $N=4$, global texture and $N=6$,
an example of `nontopological'  texture.
Of these the `generic' cases may be argued to be global strings,
which result from the complete spontaneous breaking of
any abelian continuous global symmetry, and texture,
which results from the complete breaking of any
nonabelian continuous global symmetry.
 It is
important to keep in mind, however,  that `realistic'
unified theories may
include more complex symmetry breaking schemes,
in which results different by a factor of two might well
be obtained.
For example the family symmetry scheme studied in \JT\
corresponds to two $N=6$ theories, coupled in
such a way as to produce topological texture.

In previous work, Bouchet, Bennett
and Stebbins \ref\BBS{F. Bouchet, D. P. Bennett and A. Stebbins,
Nature {\bf 335} (1988) 410.}
constructed an anisotropy map for  gauged
cosmic strings.
They used a flat spacetime approximation to compute the anisotropy
and ignored electron scattering, which in a reionised universe
acts to smear out the distinctive linelike discontinuities
produced by cosmic strings \ref\KS{N. Kaiser and A. Stebbins,
{\it Nature}, {\bf 310} (1984) 391.}, for redshifts earlier
than last scattering. Reionisation seems
likely in the string scenario:
in this case the strings would only start to
produce clear discontinuities well after last scattering, on scales
in excess of $\theta_{ls}\sim 6^o$. And of course these discontinuities would
be very heavily masked by the smaller angular scale fluctuations
produced at earlier epochs.

In our calculation for $N=2$, we do not aim at infinite resolution,
but rather to produce a map smoothed on a scale comparable
to the beam size adopted for degree scale measurements. We
include the effects of electron scattering ignored in the earlier
calculations, and we
believe that
this provides a more realistic picture
of the anisotropy pattern one could expect from
a degree scale experiment.

\centerline{\bf Techniques}

We study the
evolution of the symmetry breaking scalar fields
using, for $N>2$,  the finite difference scheme of \PST\ for the nonlinear
sigma model.
We are able, with these techniques, to evolve the scalar fields
in boxes in excess
of $200^3$ grid points, and for $N>2$ to
convincingly demonstrate that
the
results are independent of box size.
For $N=2$,
we have developed
a more accurate difference scheme to evolve a single
`angular' field. In this case we do find a systematic dependence
on box size (results change by of order 20 \% when we double the box size)
so the results should be treated with some caution. For this case
we normalised to COBE and computed the degree scale
anisotropy using exactly the same physical resolution.

In this paper we extend the techniques developed in \PST\ for calculating
the large angular scale anisotropy, including the
effects due to  photons scattering off electrons in the ionised medium.
The effect of photon drag on the velocity of the ionised electrons
(and baryons)
is a strong function of redshift, the ratio of the
collisional damping rate $t_d^{-1}$ to the expansion rate $t^{-1}$
being given by
$t/t_{d} \sim \bigl[(1+Z)/200\bigr]^{5\over 2} h^{-1}$.
Since the redshift of last scattering is
somewhat smaller than 200
in the theories we are interested in, we make the
approximation of  ignoring  photon drag, so that the velocity of
the scattering electrons simply equals the local
velocity of the dark matter. This simplifies the calculation
enormously - we do not need to evolve the baryon-electron-photon
fluid. A useful check on this approximation is that for
scale invariant adiabatic initial conditions
we should reproduce the results for an (unphysical) standard CDM model
without recombination. We have found reasonable agreement
for the spectrum of multipoles obtained using
our technique and that of
Crittenden et. al. \ref\CD{R. Crittenden and
R. Davis, private communication.} obtained using the Boltzmann
equation
\ref\cgpt{
D. Coulson, P. Graham, U. Pen, and N. Turok, in preparation, 1993.}.

Furthermore,
for reasonable values of the current cosmological parameters,
at the relevant epochs the universe was matter dominated and
close to critical density.
This means that effectively only one parameter
enters our calculations - the horizon scale when
optical depth was unity.
One can simply
rescale the angular size of our maps to obtain results
for other values of the cosmological parameters $\Omega$,
$\Omega_B$, $\Lambda$, and the ionisation fraction $f$.

As in  \PST,
we solve the
linearised Einstein equations analytically by  Fourier
transforming them
and decomposing them into scalar, vector and tensor components.
The relevant
metric perturbations are obtained as analytic integrals
of the appropriate pieces of the scalar field stress tensor,
 computed using  FFT's at each timestep.
We perform a Monte Carlo
calculation of the temperature anisotropy
by simulating photon trajectories and computing
the energy shift each photon experiences directly.
One must choose  a large enough number of photons
so that the `shot noise' due to random scatterings,
particularly the Doppler term from late scatterings,
is averaged out on the angular scales of interest.
In practice we have found that for
a few hundred photons along
each line of sight was sufficient to suppress the white noise
component in the multipole spectra down to angular scales
correponding to one or two grid spacings.
A typical run employs a box of $100^3$ for the field evolution,
$50^3$ for the metric perturbations and photon trajectories,
with 200 photons along each line of sight to make a $50^2$ map.
It takes around 5 hours on the Convex C-220 at Princeton.
Our technique
may be readily extended to compute polarisation effects.

In synchronous gauge the temperature anisotropy
experienced by photons along any path segment
is given to first order by
the Sachs Wolfe formula
\eqn\dl{\eqalign{
&\qquad \qquad {\delta T \over T}({\bf n}) =
-{1\over 2} \int_i^f d\eta h_{ij,0}(\eta, {\bf n}\eta) n^i n^j}}
where ${\bf n}$ is the direction vector of the photon,
 $h_{ij}(\eta, {\bf x})$ is the perturbation in the spatial metric,
and $\eta$ is conformal time ($h_{0\mu}=0$ in synchronous gauge).
In synchronous gauge the velocity on the dark matter is zero
at all times (in this linear regime of course), and there is
no explicit Doppler term. However,
after integration by parts  the
expression
may be decomposed into the following pieces
\eqn\dla{\eqalign{
{\delta T \over T}({\bf n}) &=
-{1\over 2} \int_i^f d\eta (h_{ij,0}^S +h_{ij,0}^V +h_{ij,0}^T)
(\eta, {\bf n}\eta) n^i n^j
+\bigl[ - {\bf n}. {\bf \nabla} \chi(\eta, {\bf x})\bigr]^f_i \
}}
where the surface term represents the Doppler effect of
the photon scattering off moving electrons.
This last term
was not included in our previous work
\PST, as it is unimportant on angular scales larger than the horizon
at last scattering.
In Fourier
space the expression for the velocity potential $\chi$ is
\eqn\dlaa{\eqalign{
\chi(\eta,{\bf k}) =& 8 \pi G\int_0^\eta d\eta' \bigl( {1\over 45} \eta \eta'
- {1\over 30} \eta^{-4} \eta'^6 -{1\over 9} \eta^{-2} \eta'^4 \bigr)
(\Theta +2 \Theta_S)(\eta',{\bf k})\cr
}}
where $\Theta$, and $\Theta_S$ are scalar components of the
source fields stress tensor. Explicit expressions for the
$h_{ij,0}^{S,V,T}$ are given in \PST.

The initial conditions in
these theories are taken to be  zero temperature
distortion and zero metric perturbation. Each of
the physical quantities
of interest in the present calculation - the path dependent
gravitational redshift term, the Doppler term, and the
`intrinsic' anisotropy term, can be shown to be dominated
at the appropriate horizon crossing epoch. Thus the calculation
is insensitive to the early  behaviour of the source
fields, and issues of `compensation' \ref\SV{S. Veeraghavan and A.
Stebbins, Astrophysical Journal {\bf 365} (1990) 37.}
do not arise.
To construct photon trajectories we evolve each photon backwards
in time along
paths starting at the position of the observer. We compute
the probability of a scattering event occurring during each timestep,
and then decide whether one actually occurred by calling
a random number. The information needed to reconstruct
the entire  photon trajectory (initial position and direction, plus
an appropriate random number seed) are stored and later used to step
along the photon paths during a simulation.

For the earliest part of its trajectory, a photon executes a random
walk with step length much less than the horizon scale.
During this early phase,
under the assumption that
the metric perturbation is constant over the scales traversed
by the photon, it can be seen from \dl\
that the photon aquires an energy shift of just
$-{1\over 6} h$. We use this as an initial condition at the
start
of each photon trajectory,
at the time when the photon mean free path equals one grid unit.
We compute the scalar `velocity potential'
$\chi$  and the metric perturbations $h_{ij,0}^{S,V,T}$ in \dla\
in real space as the fields evolve, and use them to determine the
energy shift experienced along each photon path segment,
and at each scattering event.

\centerline{\bf Results}

Maps of the CMB anisotropy produced by global strings ($N=2$),
monopoles ($N=3$), textures ($N=4$) and a representative
case of `nontopological texture' ($N=6$) are shown in
Figure 1 a-d. Figure 1e shows an equivalent map
for the standard CDM theory \ref\BE{G.P. Efstathiou and J.R. Bond, private
communication}. The maps have been smoothed
with a Gaussian window of FWHM $1.5^o$,
and the angular scale subtended by the box side is
$\Theta=30^o$  and $20^o$
for the field ordering theories and standard CDM repectively.
These angular scales are for
$h=0.5, \Omega_B= 0.0125$, $f=1$, $\Omega=1$, $\Lambda=0$.
For other values of the cosmological parameters $\Omega$, $h$,
$\Omega_B$, $\Lambda$, and the ionisation fraction $f$
the maps can be simply rescaled, for example if $\Omega=1$, $\Lambda=0$
the angular sizes
scale as $\theta \propto (\Omega_B f h_{50} /0.05)^{1\over 3}$.

In Figure 2, we show the  multipole spectra
for
$N=2,3,4$ and 6, computed by
averaging over
the six maps obtained by observers looking at
each face of a cube.
At small $l$ the results are consistent with previous
all sky simulations \PST.
There is significant scatter (`cosmic variance'), particularly
at lower $l$, which probably accounts for all the
apparent `features' on the curves - none appear statistically
significant.
There is a striking suppression of power at higher $l$,
most significant in the texture and
nontopological texture theories. {\it At $l\sim 100$,
there is ten times less power in the texture theory than in standard CDM,
if both are normalised to COBE}. Note however that this
holds only if $\Omega=1$, $\Lambda=0$. In an open universe, or
in a $\Lambda $ dominated flat universe, there would be suppression
of the anisotropy on COBE scales,  requiring a larger value
for $\epsilon$, and increasing the power on degree scales.

Is the suppression entirely a result of reionisation? We have compared
these spectra with those produced by standard CDM with complete reionisation
assumed,
 calculated both using our
techniques and also from the Boltzmann equation \CD. For that case, $l^2 c_l$
is approximately flat, but with a small Doppler bump,
 up to around $l\sim 70$, before falling off
to very small values for $l>100$. The lack of high $l$
power in the texture and nontopological texture cases
seems to be a result both of reionisation, and the fact that the
anisotropy comes mainly from a time dependent potential with
a large coherence length.

In spite of the strong suppression of higher $l$ moments,
there is {\it some}  power up to very large $l$.
Some of this high $l$ power is undoubtedly attributable to
the localised defects themselves, but the best way to look for
defects is certainly not in the power spectrum.

\vfill\eject
\centerline{\bf NonGaussianity}

The simplest measures of  nonGaussianity
are  the third and fourth irreducible moments of
the temperature distribution, known in dimensionless form as
the skewness and kurtosis. Using the six maps produced
for each theory, we have calculated the mean skewness and
kurtosis. None of these appeared to be statistically significant - i.e.
the mean never deviated by many standard deviations from zero.

Another crude test is the level of
maxima and minima in the maps compared to the r.m.s. value.
The monopole maps
show strong positive and negative peaks at a level of
$\pm 3.5 \sigma$. The texture maps also appear to be significantly
nonGaussian - the cold spot in Figure 1c is a $3.7 \sigma$
event, and other maps showed similar hot and cold spots.
It is worth emphasising that strong cold spots with
peak level around $(\delta T/T)\sim  5 \times 10^{-5}$
are expected in the texture theory, and
are hard to  mimic with astrophysical sources.
The $N=6$ maps appear more Gaussian, in line with
expectation, and
have maxima and
minima at around the $2.5 \sigma$ level, similar to
the standard CDM map. These numbers are suggestive, but
a more detailed assessment of the
level of nonGaussianity requires many more realisations.
We have applied an optimised statistic for
detecting nonGaussianity \ref\gt{P. Graham and N. Turok,
in preparation (1993).} to scans simulating the
Schuster et. al. experiment on our maps, and  do find
the nonGaussianity to be detectable with signal to noise
ratios as low as 1.5.
It is clear however that
detecting the nonGaussianity
requires a sophisticated statistical treatment,
as well as very clean experimental
data.

\centerline{\bf Conclusions}

We have developed the techniques required for a full relativistic
computation of the degree scale CMB anisotropy in nonGaussian
theories of structure formation based on symmetry breaking and phase
ordering. It is now straightforward to construct a statistical
ensemble of microwave sky maps from which results for any experiment
may be predicted, and confidence levels deduced.
Our main conclusion is that the goal of distinguishing
theories based on
field ordering from Gaussian theories
based on inflation should be readily achievable from microwave
anisotropy
data available
in the near future. One  striking difference is in the spectra
of multipoles, caused partly by the reionisation expected in
theories with topological defects. But even more distinctive
is the nonGaussianity of the anisotropy pattern
- which may be detectable by the
degree scale experiments currently in progress.

\centerline{\bf Acknowledgements}
We would like to thank P.J.E. Peebles and D. Spergel
for useful discussions. DC thanks the SERC (UK) for financial
support.
The work of NT was partially  supported by the SERC (UK),
NSF contract PHY90-21984,  the Alfred P. Sloan Foundation, and
the David and Lucile Packard Foundation.

\baselineskip=12pt
\listrefs

\def\doublespace{\baselineskip=20pt}
\def\singlespace{\baselineskip=10pt}
\def\lbaselines{\baselineskip=10.10pt   
                     \lineskip=0pt
                     \lineskiplimit=0pt}
\def\oneskip{\vskip\baselineskip}         
\def\blankline{\oneskip}
\hoffset=1.92 truecm    
\voffset=1.2  truecm    
\hsize=15 truecm      
\vsize=22 truecm      
\nopagenumbers

\baselineskip=16pt

\centerline{\bf Table 1. Standard Deviation on $1.5^o$ scale}
$$
\vbox{
\halign{# \hfil & \quad # \hfil
&  \hfil # & \quad \hfil #  \cr
\noalign{\hrule}
\noalign{\smallskip}
\noalign{\hrule}
\noalign{\medskip}
{Defect} & {N } & { $\delta T/T)_{1.5^o}$  } \cr
\noalign{\medskip}
\noalign{\hrule}
\noalign{\medskip}
\cr
Strings & 2 & $ 1.8\pm 1.0 \times 10^{-5 }$\cr
Monopoles &3 & $1.7\pm 0.4\times 10^{-5}$\cr
Textures &4  & $1.5\pm 0.3 \times 10^{-5}$\cr
N.T. Textures & 6 & $1.0 \pm 0.2\times 10^{-5}$\cr
Standard CDM & & $3.0  \times 10^{-5}$\cr
\cr
\noalign{\hrule}
\noalign{\smallskip}
\noalign{\hrule}
\noalign{\smallskip}}}$$
$$\vbox{
\noindent
Table 1. The standard deviation of the anisotropy maps
after smoothing with a Gaussian of FWHM = $1.5^o$. All theories
have been normalised to fit the standard deviation on
the 10$^o$ scale reported by COBE, $(\delta T/T)_{10^o}= 1.1 \pm
0.18 \times 10^{-5}$.
The errors quoted on the field ordering theories
are estimates of our (one sigma equivalent) systematic errors.
Statistical errors are of order 10\%.
The number for standard CDM is the value for the map shown in Figure 1e
\BE.
 }
$$

\baselineskip=12pt \vfill
\centerline{\bf Figure Captions}

\fig{1}{ Temperature anisotropy maps in the global
string, monopole, texture, nontopological texture
and standard CDM
scenarios.  The side of the box subtends an angular scale
of $30^o$ for the field ordering cases, $20^o$ for standard CDM.
$\delta T/T$ is given in units of $\epsilon =8 \pi^2 G \phi_0^2$ for
the field ordering theories for
$N>2$, where $\phi_0$ is the symmetry breaking scale,
and $8 \pi G \mu$ for $N=2$, where $\mu$ is the
string tension. Normalising to COBE,
$\epsilon= 5.9 \pm 1.1 \times 10^{-5}, 1.0\pm 0.2 \times 10^{-4}$
and $1.6 \pm 0.3 \times 10^{-4}$ for $N=3,4$ and $6$ respectively,
while $8\pi G \mu = 5.8 \pm 2.9 \times 10^{-5}$ for $N=2$\ref\bug{This value
was obtained using an improved code, and is slightly lower
than that given in \PST.}.
All the maps
have been smoothed with a Gaussian of FWHM $1.5^o$,
similar to the beam for the South Pole experiments of refs.
\Gai,\Sch.
}

\fig{2}{ Power spectra for the degree scale
temperature anisotropy in the global
string, monopole, texture, nontopological texture
and standard CDM
scenarios.  If the temperature anisotropy is
expanded as a sum of spherical harmonics, $(\delta T/T)=
\Sigma a_{lm} Y_{lm}(\theta ,\phi )$, then
$c_l$ is defined as the ensemble average $<|a_{lm}|^2>$. The power
per logarithmic interval in $l$ is then $l^2 c_l$, constant
for a scale invariant pattern.}

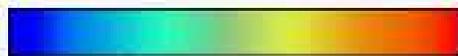
\bye